\documentclass{article}
\usepackage{graphicx} 
\usepackage{hyperref}
\usepackage[backend=biber, style=ieee]{biblatex}
\usepackage{booktabs}
\usepackage[T1]{fontenc}
\usepackage{pdflscape}

\title{Mapping the Landscape of Open Access Dashboards – A Dataset for Research and Infrastructure Development}
\author {Johannes Schneider}

\author{
Johannes Schneider\thanks{Helmholtz-Association, Helmholtz Open Science Office, Telegrafenberg, 14473 Potsdam, Germany.
Email: \href{mailto:johannes.Schneider@os.helmholtz.de}{johannes.schneider@os.helmholtz.de}. \textbf{ORCID}: \href{https://orcid.org/0009-0004-6510-166X}{0009-0004-6510-166X}}\\
{\small Helmholtz}
\and 
Heinz Pampel\thanks{Humboldt-Universität zu Berlin, Berlin School of Library and Information Science (IBI), Dorotheenstr. 26, 10117 Berlin, Germany and Helmholtz-Association, Helmholtz Open Science Office, Telegrafenberg, 14473 Potsdam, Germany.
Email: \href{mailto:heinz.pampel@hu-berlin.de}{heinz.pampel@hu-berlin.de}. \textbf{ORCID}: \href{https://orcid.org/0000-0003-3334-2771}{0000-0003-3334-2771}}\\
{\small Humboldt \& Helmholtz}
}

\date{May 2025}

\begin{document}

\maketitle 

\begin{abstract}

As Open Access continues to gain importance in science policy, understanding the proportion of Open Access publications relative to the total research output of research-performing organizations, individual countries, or even globally has become increasingly relevant. In response, dashboards are being developed to capture and communicate progress in this area. 
To provide an overview of these dashboards and their characteristics, an extensive survey was conducted, resulting in the identification of nearly 60 dashboards. To support a detailed and structured description, a dedicated metadata schema was developed, and the identified dashboards were systematically indexed accordingly. To foster community engagement and ensure ongoing development, a participatory process was launched, allowing interested stakeholders to contribute to the dataset.
The dataset is particularly relevant for researchers in Library and Information Science (LIS) and Science and Technology Studies (STS), supporting both empirical analyses of Open Access and the methodological refinement of indicators and policy instruments in the context of Open Science.

\end{abstract}

\section{Background \& Summary} 

The term Open Access refers to making scholarly publications openly accessible and reusable on the internet without financial, technical, or legal barriers. Open Access is shaped by the so-called BBB definition. At three conferences — in Budapest (2001), Bethesda (2003), and Berlin (2003) — new approaches to digital scholarly communication were discussed in the early 2000s, which led to a shared understanding of the term Open Access within the scholarly community. This understanding is summarized under the term BBB definition of Open Access (\cite{suber_open_2012}).

Building on numerous initiatives by researchers, Open Access has been established as a science policy objective in many countries. Key actors in this context are research funders who have embedded Open Access mandates into their funding programs. A role model for integrating Open Access into research funding is the ``Public Access Policy" of the National Institutes of Health (NIH) in the United States, which was established in 2008  (\cite{national_institutes_of_health_analysis_2008}) and updated in 2025 (\cite{Lenharo2025}). The European Commission established a similar requirement in its Horizon 2020 research framework program, which has been in place since 2014 (\cite{european_commission_access_2015}).

Individual countries and alliances of countries have also adopted Open Access as a policy goal. One example is the European Union, which already set the goal in 2012 that by 2016, 60 percent of publicly funded research results in Europe should be freely accessible online (\cite{european_commission_background_2016}). Another target was set in 2016, when the EU Council called for a ``full scale transition towards open access" by 2020 (\cite{council_of_the_european_union_council_2016}).

In recent years, various stakeholders have developed dashboards to track Open Access. In the field of information management, dashboards are understood as tools for visualizing data. Such tools, which have their roots in the Executive Information Systems (EISs) of the 1980s (\cite{Few2006}), gained significant relevance during the COVID pandemic, when real-time communication of pandemic-related data became essential (\cite{dong_johns_2022}; \cite{khodaveisi_characteristics_2024}; \cite{vahedi_applications_2022}). 
These dashboards enable the communication of the current state of affairs using defined indicators. \textcite{Few2006} defines dashboards as follows: ``A dashboard is a visual display of the most important information needed to achieve one or more objectives; consolidated and arranged on a single screen so the information can be monitored at a glance."

With the growing prominence of Open Access, dashboards have increasingly been applied in scholarly information management. They are often built on data infrastructures that fall under the category of research information, referring to data on scholarly activities and outputs. Many academic libraries operate specialized Current Research Information Systems (CRIS) to collect and manage such data. Over the past decade, these systems have increasingly included information on the access status of scholarly publications, such as journal articles (\cite{biesenbender_using_2019}).

The collection of research information in general and publication output in particular can be of value to trace the Open Access transformation. For example, collecting data on Open Access publications can be used in these four areas:  

(1) Tracking Institutional Progress: To establish effective Open Access indicators within institutional strategies, institutions must actively monitor their publication output. This includes understanding the proportion of Open Access publications in relation to the total scholarly output of a given institution or unit, such as a research performing organization (RPO) or a corpus of publications funded by a research funding organization (RFO). Such insight is essential for evaluating the effectiveness of measures and tracking progress toward Open Access goals.

(2) Compliance with Funding Requirements: To ensure compliance with Open Access mandates set by RFOs, research institutions have a strong interest in establishing workflows that guarantee adherence to these requirements. Current research information systems (CRIS) can support these efforts by tracking publication metadata and license information.

(3) Publisher Agreements: To support decisions about entering into or discontinuing publisher agreements, such as transformative agreements, RPOs are increasingly relying on their own analyses rather than on usage statistics provided by publishers, in order to assess the cost-effectiveness of subscription-based and Open Access publishing models. This shift is reshaping collection development practices in academic libraries. Comprehensive data on publication output and Open Access status, combined with institution-specific analyses, is essential for evaluating the financial implications of Open Access publishing.

(4) Research: To advance bibliometric and science policy research, data on the Open Access status of publications are critical for analyzing the dissemination of knowledge across countries, disciplines, and institutions. Studies in the fields of LIS and STS, such as \cite{archambault_proportion_2014} and \cite{Piwowar2018}, have demonstrated the value of such data for evaluating Open Access trends in scholarly communication.

The use of dashboards to meet information needs in the area of Open Access is gaining increasing importance. \citeauthor{salamoura_challenges_2024} published a noteworthy article on Open Access monitoring, which focuses on the landscape of existing dashboards. They highlight the different levels of monitoring and distinguish between international, national, and institutional approaches to tracking Open Access. 

The following paragraphs outline key developments in Open Access monitoring at the international, national, and institutional levels, as well as in the emergence of related dashboards.

International level: 
A look at the development of dashboards at the international level shows that the launch of the EU Commission’s Open Science Monitor in 2017 marked a significant development. This service was implemented with input from RAND Europe, Deloitte, Digital Science, Altmetric, and Figshare (\cite{rand_monitoring_nodate}). In 2018, the monitor was revised with the involvement of Elsevier. The publisher’s participation sparked widespread discussion at the time (\cite{moody_elsevier_2018}, \cite{moody_hated_2018}) and was also met with methodological criticism (\cite{french_open_science_committee_feedback_nodate}). As of May 2025, the monitor is no longer available online. Only information related to Open Research Data remains accessible on the European Commission’s website (\cite{european_commission_facts_nodate}). In 2020, the OpenAIRE Monitor was launched as part of the EU-funded OpenAIRE project, and in 2022 it was expanded to include institutional dashboards (\cite{pispiringas_new_2022}). The efforts to visualize data on Open Access continued in subsequent years under the name Open Science Observatory. These dashboards visualize data from the OpenAIRE Research Graph. Further development is planned within the framework of the European Open Science Cloud, as part of the EOSC Future project. (\cite{oneill_introducing_2022}).

National level:
Initial considerations for national Open Access monitoring were already discussed in Finland in 2017 (\cite{olsbo_measurement_2017}). The first fully implemented nationally operated Open Access dashboard was likely introduced in Germany with the Open Access Monitor (OAM, \cite{mittermaier_rolle_2021}). The development of the OAM began in 2018 as part of the project ``Synergies for Open Access – Open Access Monitoring (SynOA),” which was funded by the German Federal Ministry of Education and Research (BMBF) (\cite{barbers_synoa_2021}).
In France, the Baromètre français de la Science Ouverte was launched in 2019 (\cite{bracco_promoting_2022}). Switzerland followed with the release of its Open Access dashboard in 2022 (\cite{lib4ri_swiss_2022}). In Austria, national coordination processes for Open Access monitoring were initiated in the context of the project Austrian Transition to Open Access (AT2OA) (\cite{danowski_report_2018}). In 2024, the ``Open Access Monitor Austria" went online with a beta version.  

Institutional level:
In the field of institutional Open Access monitoring approaches, a distinction can be made between the activities of (1) RPOs and (2) RFOs.

(1) RPOs: The Forschungszentrum Jülich in Germany published its first Open Access Barometer in 2016. This dashboard was developed to visualize the transition from subscription-based journals to Open Access. The barometer provides detailed insights into expenditures for scholarly publications, particularly regarding article processing charges (APCs), and tracks the development of spending categories over time. It serves as a central monitoring tool supporting the institution’s Open Access strategy (\cite{forschungszentrum_julich_open-access-barometer_nodate}).
Another example of an institutional dashboard is the Open Access Monitor of the University of Zurich, which went online in 2021 (\cite{universitat_zurich_check_2024}).
In the context of German universities, another approach operates at the departmental level within a university. For example, the Department of Earth Sciences at Freie Universität Berlin has implemented a specific Open Science Dashboard that visualizes information on Open Access based on the department's members (\cite{duine_initiating_2024}).

(2) RFOs: At the level of funding organizations, notable examples include the Clearinghouse for the Open Research of the United States (CHORUS) (\cite{dylla_chorus_2014}), the Europe PMC Funders’ Dashboard (\cite{europe_pmc_team_europe_2020}), as well as the dedicated dashboards for funders provided within OpenAIRE (\cite{brunschweiger_new_2023}). It should be noted, however, that several of these solutions are not operated directly by the funding organizations themselves. For example, CHORUS is an initiative led by publishers (\cite{poynder_open_2013}).

So far, there are few established recommendations for Open Access monitoring. 
In Austria, the following ten recommendations for Open Access monitoring were developed as part of the AT2OA project. These serve as guiding principles: ``1. Develop a common framework for the definition of Open Access categories, 2. differentiate between monitoring publication output (OA share) and monitoring Open Access publishing costs, 3. ensure sustainability, 4. take existing infrastructures into account, 5. Open Access monitoring should be open and transparent, 6. Open Access monitoring must provide added value for the participating institutions, 7. implement a shared monitoring tool, 8. identify data providers, 9. define minimum standards and interfaces using OAI-PMH, 10. `keep it simple' – start small and add complexity later." (\cite{danowski_empfehlung_2020})

As part of the collaboration among RPOs and RFOs within Science Europe, a set of guidelines for Open Access monitoring was developed. Starting from the questions of why and what to measure, the guidelines outline procedures for the collection of data on Open Access and offer recommendations for gathering and interpreting publication data. They address key aspects such as the use of persistent identifiers, data sources, data aggregation, as well as the interpretation and reporting of results  (\cite{philipp_open_2021}). 

As outlined above, a growing number of dashboards have been developed in recent years to monitor various aspects of Open Access. However, a comprehensive overview of these tools has been lacking. The project OA Datenpraxis (Data practice for shaping the Open Access transformation - analysis, recommendation, training \& networking), funded by the German Research Foundation (DFG), responded to this need by systematically mapping and analyzing the current landscape of Open Access dashboards (\cite{pampel_datenpraxis_2024}). 

As a result, we compiled a collection of Open Access dashboards. The procedures for data collection and metadata annotation, based on a custom-designed schema, are described in detail in the Methods section. As of May 2025, our dataset included 58 dashboards. 

All dashboards in the collection contain data on textual publications. 34 of the listed dashboards also contain data on different scientific outputs, most notably on research data or research software. There are a few dashboards with more specialized categories, such as a monitor of clinical trials in the French Open Science Monitor or data on open metadata in the Hybrid Open Access Dashboard.

11 dashboards present data on scientific output from research institutions, 29 dashboards have data on national scientific output and 18 dashboards have data on an international level. All institutional dashboards are European, 15 national dashboards are European, with the remaining dashboards all from the CHORUS collection (12 dashboards for US agencies, one Australian, one Japanese dashboard). 6 international dashboards are also European, whereas the remaining 12 international dashboards are all global.

\section{Methods}

\subsection{Data Search}

The largest share of dashboards was found via online search during the time period from 2024-10 to 2024-12 with a few additions until 2025-05. Dashboards were identified by searching for combinations and variations of the search terms `open access', `open science' and `dashboard', `monitor' in two search engines (DuckDuckGo and Google). Whether the dashboard was still actively maintained was not considered an inclusion criterion, only its accessibility. The relevant criterion for inclusion was whether the dashboard contained data on some aspect of Open Access of scientific output, be that textual publications, research software, research data or other forms.  In line with a broad definition of Open Access dashboards, we included not only platforms that display Open Access shares for specific entities, but also services like the CWTS Leiden Ranking, insofar as they contain relevant Open Access data, even if this is not their main objective. Properties of the identified dashboards were gathered in an Excel table based on the metadata schema described below.

In two cases, we contacted individuals involved in a dashboard that was not available: in one case, an existing dashboard with a broken link (Hamburg Open Science Dashboard, no reply), and a not yet existing dashboard (The Biomedical Research Open Science Dashboard project, \cite{cobey_open_2021}\footnote{https://osf.io/jm8wg/}; it will be added to the overview as soon as it becomes available). One dashboard was removed during research since the dashboard itself was removed or is at least not locatable any more (Open Science Monitor by the European Commission\interfootnotelinepenalty10000\footnote{previously available with graphics on Open Access under \url{https://research-and-innovation.ec.europa.eu/strategy/strategy-research-and-innovation/our-digital-future/open-science/open-science-monitor_en}}). In one case, the website of a dashboard is still available, but the links to the dashboards of the specific years are not functional any more.\footnote{Dutch Open Science Dashboard by DANS; \url{https://dans.knaw.nl/en/data-expertise/monitoring-and-analysis/}, superseded by two other Dutch dashboards}

There are two dashboard collections heavily represented in the overview: the CHORUS Dashboards and OpenAIRE Monitor Dashboards. The CHORUS dashboard collection contains 14 dashboards, on various US agencies and departments, one Australian and one Japanese agency. The OpenAIRE Monitor Dashboard collection contains 97 dashboards, of which only 14 are publicly available and two of these are empty dashboards (all data as of May 2025). The French Open Science Monitor contains links to over 50 local institutional sub-versions of itself\footnote{\url{https://frenchopensciencemonitor.esr.gouv.fr/declinaisons/bso-locaux}}, some of which are not reachable. We have decided not to list these sub-variants of the collective version of the French Open Science Monitor in order to avoid a bloating of the overview.

\subsection{Development of the Metadata Schema}

The metadata schema (see Table 1) was developed to systematically describe properties of Open Access dashboards. The process began with an analysis of existing metadata standards and best practices in Open Science, ensuring compatibility and interoperability with widely used frameworks. Key elements were identified based on the characteristics and functionalities of the dashboards identified in the search process, including aspects such as data sources, visualization methods, accessibility, and update frequency.

The schema was iteratively refined through expert feedback and community input, ensuring that it captures essential attributes while remaining flexible enough to accommodate new developments. By structuring the metadata in a standardized format, the schema facilitates comparability, discoverability, and integration of Open Science dashboards into broader research infrastructures. The description was guided by established and widely recognized metadata schemas. In doing so, we were able to build on relevant prior work by members of the author team, particularly in the context of developing the metadata schema for the re3data registry of research data repositories (\cite{strecker_metadata_2023}).

Decisions for the entry for `Name', including decisions about capitalization, were determined based on multiple sources, in order of descending priority: if the dashboard website itself recommends a form of citation, or if there is an accompanying publication describing the dashboard, the version of the name used in those sources is taken as definitive; otherwise, the self-description on the website is used (as in the header, the `about' page or the self-referring term throughout the website). If there is more than one term used, we choose the more informative one. If acronyms are used in the self-description, the acronym is adopted without spelling it out. One edge case was a visualization website for the project OpenAPC (\cite{jahn_datasets_2014}): the main project site links to the visualization site under the clickable element ``APC Treemap Visualizations" while the corresponding site offers no self-description whatsoever; therefore this term was used as entry for `Name'. 
In case of a specific instance of a dashboard collection, the naming scheme is ``[Name of dashboard collection]: [Name of organization]", e.g. ``OpenAIRE Monitor Dashboard: University of Göttingen".
In case a dashboard project provides separate dashboards for each year (or a specific time period) we removed the year from the name (i.e. it is `CWTS Leiden Ranking', not `CWTS Leiden Ranking 2024').

Under `URL' we listed the URL to the homepage of the dashboard; if the listed dashboard is only one of other dashboard-unrelated services on a website,  we provide the URL leading to the webpage of the specific dashboard.

Under `Time period' we recorded the year of the earliest data presented in the dashboard until the year of the latest data. Note that this is distinct from the year of the publication of the dashboard; i.e. a year of latest data earlier than 2024 does not necessarily mean the dashboard is inactive (the Leiden Ranking 2024 e.g. refers to data from 2019\textendash2022).

Under `Operator' we recorded the name of the organization responsible for running the dashboard and under `ROR' the ROR ID of that operator as an operable URL if available (`N/A' otherwise). If e.g. a project does not have a ROR we recorded the organization that contains or funds it that does have one. This next higher organization is entered in `Operator', separated by comma from the original operator (e.g. `Centre for Science and Technology Studies (CWTS), Leiden University'). Multiple direct operators are instead separated by semicolon. Entries in `ROR' are separated by comma. The entry in `Operator type' describes the type of the organization responsible for the dashboard. Possible values are `RPO' and `RFO' (research performing/funding organization, respectively) to distinguish two main types of research-related organizations; `other' for everything else that does not fall under this umbrella. 

`Scope' records the extent of the geographic coverage of the dashboard data. The values are: `international' if data is available for more than one country, `national' if there is data for one country and `institutional' if the dashboard captures data on one or more organizations from one country but not country-wide data. The related column `Countries' lists the location of the organization(s) whose data are represented in the dashboard. Possible values are: 3-letter ISO 3166-1 alpha-3 country codes; the value `Europe' if the dashboard contains data on more than one country from the EU OR Switzerland OR Norway OR UK\footnote{All EU-dashboards so far contain data on at least one of those non-EU countries.}; the value `global' if there is data from more than one country and more than three countries outside of `Europe'\footnote{This threshold is to prevent that an EU-based dashboard such as those by the European Commission does not get labeled as `global' simply by including the US for comparison.}.

The column `Data type' describes the kind of data that the dashboard contains. Possible values are `publications' for data on textual publications, `data' for dashboards on research data, `software' for dashboards on research software, `infrastructures' for data on repositories and `other' for everything else that does not fit into any of the previous categories.

`Dashboard license' records the copyright license of the software used to create the dashboard; if the dashboard provides a generic license for its website without specifically mentioning the software or its dataset, this license is instead entered here. Allowed values are entries of copyright licenses listed in the SPDX License List\footnote{https://spdx.org/licenses/} (e.g. CC-BY-4.0, CC0-1.0).

`Data license' is the copyright license under which the data of the dashboard is published. Controlled vocabulary for entries is the same as for the column above. 

`Data source' describes the availability of the data source the dashboard is based on: ‘open’ if the data comes from an open source such as e.g. OpenAlex, ‘proprietary’ if not; both values are entered if the dashboard uses data sources of both types.

The column `Part of' captures whether a dashboard is part of a dashboard collection, so far that is the `OpenAIRE Monitor Dashboard' and the `CHORUS Dashboard' collection (`N/A' if not applicable).

Finally, `Documentation' is a free form field in which relevant links and comments are gathered, for example links to the specific part of the website where the various licenses are mentioned, DOIs to accompanying publications about the dashboard and other comments.

\section{Data Records}

A persistent version of the dataset (Schneider et al., 2025) is available via the open repository Zenodo: \url{https://doi.org/10.5281/zenodo.15593821}

The dataset is also available on the website of the OA Datenpraxis project (see Figure 1): \url{https://oa-datenpraxis.de/en/dashboards.html}
\begin{figure}
    \centering
    \includegraphics[width=1\linewidth]{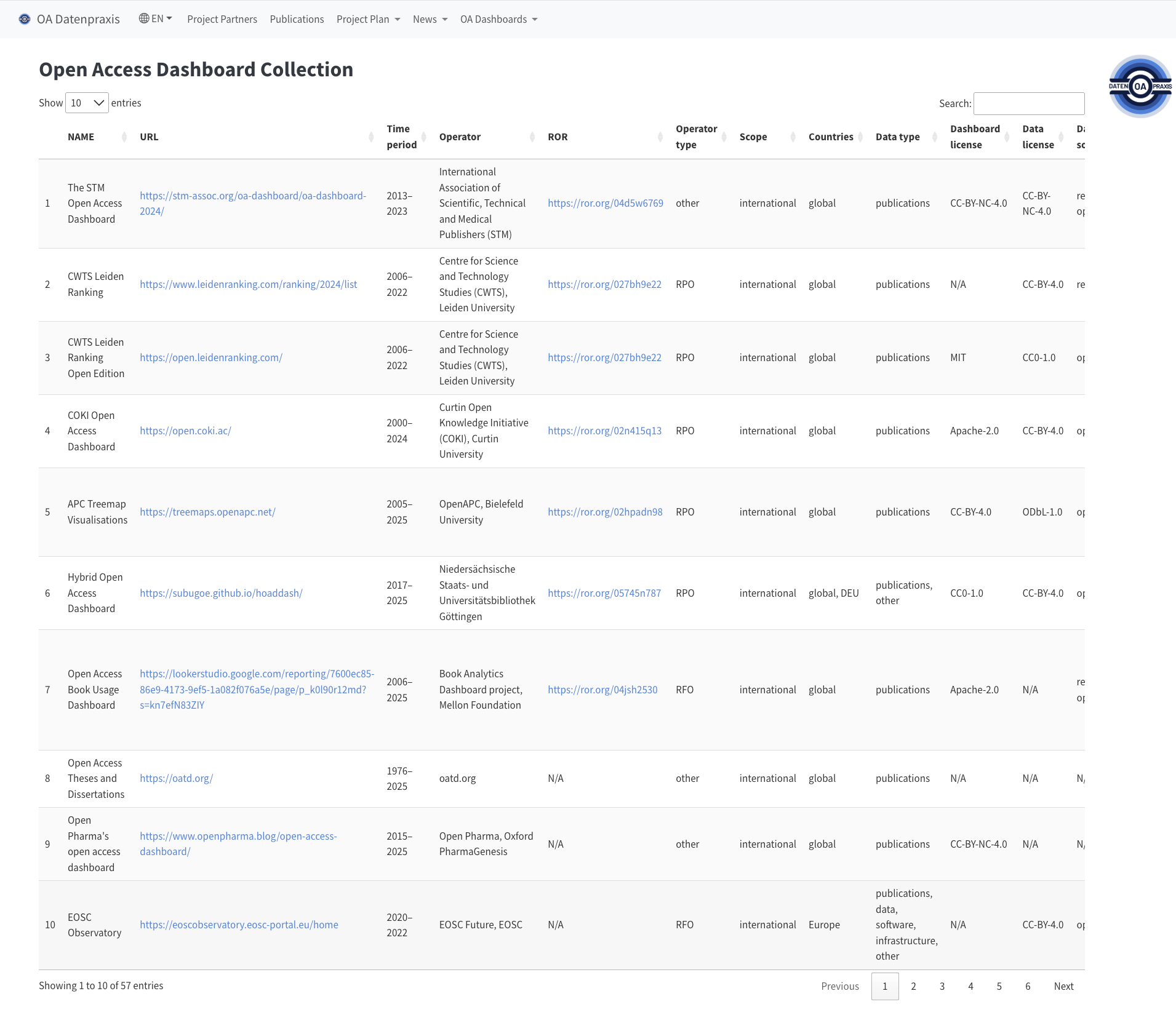}
    \caption{Dashboard website}
    \label{fig:dashboard website}
\end{figure}

There is also a version of the dataset that can be edited, for example to add new dashboards: \url{https://nextcloud.gfz.de/s/QG38J6kEDbJ8DXz}

The metadata schema is part of this data descriptor (see Table 1), and also included in the dataset (Schneider et al., 2025), and  is available on the project website: \url{https://oa-datenpraxis.de/en/dashboards_schema.html}

The dataset was compiled through a systematic and reproducible procedure aimed at ensuring consistency, reliability, and long-term usability. The data collection process, described in the Methods section, was conducted between October 2024 and May 2025 and involved a comprehensive web search using clearly defined terms and inclusion criteria. Dashboards were only included if they were accessible online and contained structured data on Open Access-related aspects of scholarly output.

Each dashboard was annotated using a custom metadata schema specifically developed for this project. The metadata schema defines 13 distinct properties, including information about dashboard scope, operator type, data type, license, and data sources. Wherever possible, we adopted controlled vocabularies to ensure interoperability with other datasets.

Metadata entries were compiled manually in a spreadsheet. Key decisions regarding naming conventions, URL selection, and attribution were standardized based on a defined hierarchy of evidence, such as citation recommendations, project documentation, or internal labeling within the dashboard interface. The metadata schema was developed with reference to existing frameworks, including the metadata schema used by the re3data registry for research data repositories \parencite{strecker_metadata_2023}.

The curation of the data according to the developed metadata schema followed a two-step process. Initially, each record was assigned a set of metadata descriptors by a first individual. This initial assignment was then independently reviewed by a second person. In cases of discrepancies, differing assessments were discussed to reach a consensus. In the few instances where divergent interpretations could not be clearly resolved, a third person was involved to adjudicate. This procedure was implemented to ensure the consistency and validity of the metadata curation.

To ensure consistency in annotation, a single curator processed the metadata entries, applying the schema across all dashboards. Cases of ambiguity, such as unclear licensing or multiple versions of a dashboard, were resolved by reviewing project documentation or contacting dashboard providers directly. Where licensing details were not publicly disclosed, these fields were marked as unknown or N/A.

The metadata schema itself is versioned and publicly accessible via the OA Datenpraxis project website. Feedback on the schema is actively sought as part of an ongoing participatory process to refine and enhance both the dataset and the underlying schema.

\section{Technical Validation}

The entries in the data set were checked by both authors and reviewed and proof-read by an external person (Jonas Höfting).

\section{Usage Notes}

This dataset provides a foundation for the systematic analysis of Open Access dashboards. It is disseminated via multiple openly accessible platforms: a persistent version is hosted on Zenodo, a curated version is available on the OA Datenpraxis project website, and an editable version is maintained in a Nextcloud instance to support community contributions. To facilitate user engagement, we have set up a dedicated contact address, and a project webpage outlines ways for interested parties to get in touch.

We actively invite researchers and practitioners in Open Science,  LIS, and STS to reuse and extend the dataset. During the collection phase, we engaged with several dashboard providers, and in June 2025, we shared the dataset and participatory approach via mailing lists and social media.

The dataset serves both practical and scholarly purposes. From an analytical perspective, it already reveals notable patterns. For instance, dashboards appear to be predominantly a European phenomenon: only 14 out of 58 exclusively cover non-European countries—all from the CHORUS Dashboard collection. While all dashboards report on textual publications, several also include research data, software, or other output types. Most dashboards cover scholarly outputs from the past two decades.

The dataset is accompanied by a dedicated metadata schema that captures essential attributes, enhancing interoperability and facilitating reuse. Further development of the schema is both feasible and encouraged. For example, additional attributes could be incorporated to reflect specific characteristics of dashboards. Such extensions can be implemented in response to evolving informational needs and would support broader reuse and community-driven refinement.

Together, the dataset and its documentation offer a basis for collaborative enhancement aligned with Open Science principles. The open availability of the collection fosters sovereign data practices in the field of Open Access, enabling the development of dashboards independently of commercial platforms.

This vision aligns with broader infrastructural developments in the research ecosystem. Open infrastructures such as OpenAlex \cite{priem_openalex_2022} exemplify the potential of transparent and reusable research metadata. Similarly, the Barcelona Declaration on Open Research Information \cite{noauthor_barcelona_nodate} highlights the importance of openness in promoting trust, accountability, and the effective reuse of research-related information. These initiatives reinforce the relevance of our dataset as a community-driven resource that contributes to a more open and interoperable scholarly communication landscape.

\section{Code Availability}

No custom code was used.

\section{References}

\printbibliography
\section{Author Contributions}

J.S. and H.P. conceptualized the study and developed the data collection methodology; J.S. curated the data; J.S. and H.P. wrote the manuscript. Both authors contributed to the data and approved the final manuscript.

\section{Competing Interests} 

The authors declare no competing interests.

\section{Acknowledgements} 

The data set was reviewed and proof-read by Jonas Höfting. Any remaining errors are ours alone.
This work has been supported by the German Research Foundation (DFG) under the project OA Datenpraxis - Data practice for shaping the Open Access transformation - analysis, recommendation, training \& networking (Grant ID: 528466070).
Heinz Pampel was partly funded by the Einstein Center Digital Future (ECDF).

\newpage

\begin{landscape}

\begin{table}[]
\begin{tabular}{llp{0.8\textwidth}lp{0.3\textwidth}p{0.2\textwidth}}
\hline
ID  & Property & Definition & Occ & Values & Example \\ \hline
1 & Name & Name given to the dashboard by its authors, from (in descending priority): provided citation form, name given in associated publication, self-description on the website (header, `about’ page, self-referring term throughout the website), browser tab name; year is removed from dashboard name if there are versions for multiple years; instances of dashboard collections are named after the scheme `[Name of dashboard collection]: [Name of organization]’ & 1 &  & French Open Science Monitor, CWTS Leiden Ranking, OpenAIRE Monitor Dashboard: Universität Göttingen \\
2 & URL & URL to homepage of the dashboard; if listed dashboard is only one of other dashboard-unrelated services on a website, URL of the specific webpage & 1 &  & \url{https://oamonitor.ch/} \\
3 & Time period & Time period in years from the year of earliest data presented in the dashboard to the year of latest data; note: this is not the year of publication in the dashboard - `year of latest data’ earlier than 2024 does not necessarily mean the dashboard is inactive; the Leiden Ranking 2024 e.g. refers to data from 2019\textendash2022 & 1-n & [year of common era]–[year of common era] & 2018\textendash2024 \\ 
4 & Operator & Name of operator responsible for running the dashboard; if that operator does not have an associated ROR, but there is an organization containing the operator that has a ROR, this organization is listed after the original operator, separated by comma; multiple direct operators are separated by semicolon. & 1-n &  & Centre for Science and Technology Studies (CWTS), Leiden University \\
4.1 & ROR & ROR of the organization(s) listed under 4 if available, N/A if unavailable & 1-n & ROR ID as an operable URL, N/A & \url{https://ror.org/027bh9e22} \\
4.2 & Operator type & Type of organization running the dashboard, distinguishing two major types of research-related organizations, Research Performing and Funding Organizations (`RPO’ and `RFO’ respectively), with `other’ for everything else & 1-n & Research performing organization (RPO), Research funding organization (RFO), other & RPO \\
5.1 & Scope & Scope of the coverage of dashboard: international if data for more than one country, national if data for one country, institutional if data on one or more organizations but not country-wide & 1 & international, national, research institution & international \\
5.2 & Countries & Country code of the organization(s) whose data are represented in the dashboard, represented with country code value `Europe’ if more than 1 country but all from EU + Switzerland OR Norway OR UK value `global’ if more than 1 country and more than 3 countries outside of `Europe’ & 1-n & Controlled vocabulary: 3-letter ISO 3166-1 alpha-3 country code; additionally: `Europe’, `global’ & DEU \\
\end{tabular}
\caption{Metadata schema}
\label{tab:metadata}
\end{table}

\begin{table}[]
\begin{tabular}{llp{0.8\textwidth}lp{0.3\textwidth}p{0.2\textwidth}}
\hline
ID  & Property & Definition & Occ & Values & Example \\ \hline
6 & Data type & Type of data that the dashboard shows `publications’ for data on textual publications such as e.g. journal articles `data’ for research data `software’ for research software `infrastructures’ for data on e.g. repositories `other’ for data that does not fall under any of the other categories & 1-n & publications, data, software, infrustructures, other & publications \\
7 & Dashboard license & Copyright license of the software used to create the dashboard; if the dashboard provides a generic license for its website without specifically mentioning the software or its dataset, this license is instead entered here & 1-n & controlled vocabulary: SPDX License List entries, N/A & MIT, Apache-2.0 \\
8 & Data license & Copyright license under which the data of the dashboard is published & 1-n & controlled vocabulary: SPDX License List entries, N/A 	 & CC-BY-4,0, CC0-1.0 \\
9 & Data source & Availability of the data source the dashboard is based on: `open’ if data comes from an open source such as e.g. OpenAlex, `proprietary’ if not, both values if dashboard uses data sources of both types & 1-n & open, restricted, N/A & open \\
10 & Collection & Names the dashboard collection the dashboard is part of if applicable; `N/A’ otherwise & 1 & string of the dashboard collection name, N/A & OpenAIRE Monitor Dashboard, CHORUS dashboard \\
11 & Documentation & Free text field in which relevant links and comments are gathered; these can include: links to the specific part of the website where the various licenses are mentioned, DOIs to accompanying publications about the dashboard and other comments & 1-n & string of comment, N/A &  	Data source: \url{https://bibliotecnica.upc.edu/en/observatori#metodologia} \\ \hline
\end{tabular}
\end{table}
\end{landscape}

\end{document}